\documentclass{article}
\usepackage{amsmath, amsthm, amssymb,amsbsy,bbm}
\usepackage{hyperref} 
\hypersetup{backref, pdfpagemode=FullScreen, colorlinks=true, linkcolor=black, urlcolor=black, citecolor=black}
\usepackage{color}
\def\bea{\begin{eqnarray}}
\def\eea{\end{eqnarray}}
\def\bean{\begin{eqnarray*}}
\def\eean{\end{eqnarray*}}

\def\bvec#1{\raise1.5ex\hbox{$\rightarrow$}\mkern-16.5mu #1}

\def\m#1{\mathcal#1}
\def\parp{\partial^+}
\newcommand{\be}{\begin{equation}}
\newcommand{\ee}{\end{equation}}

\newcommand{\barr}{\begin{array}}
\newcommand{\earr}{\end{array}}

\newcommand{\bed}{\begin{displaymath}}
\newcommand{\eed}{\end{displaymath}}
\newcommand{\bal}{\begin{array}{ll}}
\newcommand{\eal}{\end{array}}

\newcommand{\deltab}{\boldsymbol\delta}

\renewcommand{\d}{\partial}
\newcommand{\nn}{\nonumber}
\newcommand{\E}{E_{7(7)}}
\newcommand{\ka}{\kappa}
\newcommand{\ep}{\epsilon}

\numberwithin{equation}{section} 
\begin{document}
\begin{titlepage}
\begin{flushright}    UFIFT-HEP-08-02 \\ 
%hep-ph/yymmnnn 
%E7Letter-by-final.tex %<<<<<<<<<<<<<<  PLEASE WRITE THE FILENAME HERE
%\framebox{Updated: \today}
\end{flushright}
\vskip 1cm
\centerline{\LARGE{\bf {$E^{}_{7(7)}$  on the  Light Cone}}}
\vskip 1cm
\centerline{\bf Lars Brink${}^{\,a}_{}$,  Sung-Soo Kim${}^{\,b}_{}$, and Pierre Ramond${}^{\,b,\,c}_{}$}
\vskip .2cm
\centerline{\em ${}^{a~}_{}$Department of Fundamental Physics}
\centerline{\em Chalmers University
of Technology, }
\centerline{\em S-412 96 G\"oteborg, Sweden}

\vskip .2cm
\centerline{\em  ${}^{b~}_{}$Institute for Fundamental Theory,}
\centerline{\em Department of Physics, University of Florida}
\centerline{\em Gainesville FL 32611, USA}

\vskip .2cm
\centerline{\em  ${}^{c~}_{}$School of Natural Sciences, Institute for Advanced Studies,}
\centerline{\em Princeton NJ 08540 }
\vskip 1.5cm

\centerline{\bf {Abstract}}
\vskip .5cm
\noindent We use the Cremmer-Julia $\E$ non-linear symmetry of  ${\m N}=8$ Supergravity to derive its order $\kappa^2$ on-shell Hamiltonian in terms of one chiral light-cone superfield.  By requiring that $\E$ commute with the super-Poincar\'e group, we deduce to lowest non-trivial order in $\kappa$,  the light cone $\E$ transformations of all fields of the theory, including the graviton.  We then derive the dynamical supersymmetry transformation to order $\kappa^2$, and express the Hamiltonian as a quadratic form in the chiral superfield.  
\vfill
\begin{flushleft}
January 2008 \\
\end{flushleft}
\end{titlepage}

%%%%%%%%%%%%%%%%%%%%%%%%%%%%%%%%%%%%%%%%%%%%%
%%%%%%%%%%%%%%%%%%%%%%%%%%%%%%%%%%%%%%%%%%%%%
\section{Introduction}
The maximally supersymmetric ${\cal N}=8$ Supergravity \cite{Cremmer:1979up} and ${\cal N}=4$ SuperYang-Mills \cite{BSS} play a very important r\^ole in modern theory. In the standard descriptions they look quite different and are naturally related to eleven- and ten-dimensional theories, respectively. In the light-cone frame description, however, they are described in a remarkably similar way hinting at a deep relation between them. In four dimension,  they are the only two (except possibly for some higher-spin theories) that are  described by one chiral {\em constrained} light-cone superfield which captures {\em all} their physical degrees of freedom \cite{Brink:1982pd}.  Also,  tree level Supergravity amplitudes are related to the square of Yang-Mills amplitudes \cite{Berends:1988zp, Kawai:1985xq}, and the light-cone Hamiltonian of both theories can be written as a {\em positive definite} quadratic form in their superfields \cite{Ananth:2005zd,Ananth:2006fh}. Some even suggest that the ultraviolet finiteness of  ${\cal N}=4$ SuperYang-Mills \cite{UVfinite} might extend to ${\cal N}=8$ Supergravity \cite{Bern:1998ug,Green:2006gt}. There are important structural differences though.  ${\cal N}=8$ Supergravity, unlike ${\cal N}=4$ SuperYang-Mills, is not Superconformal invariant. Instead it has  the on-shell, non-linear Cremmer-Julia,  $\E$ duality symmetry \cite{Cremmer:1979up}.  {It is therefore natural to ask if this symmetry can be exploited to bring simplicty to the quartic and higher-order interactions of $\m N=8$ Supergravity.  In this letter, as a first step in this direction,} we show how to exploit this symmetry to construct the light-cone Hamiltonian to order $\kappa^2$. Our resulting expression is remarkably simpler than a recent formulation of the same Hamiltonian with over ninety terms \cite{Ananth:2006fh}.  In this process we will also get the $\E$ transformations to lowest order for all the fields in the theory. The details of the calculations will be presented elsewhere \cite{BKR}.

After a brief review of $\E$ duality in the covariant formalism with the scalar and field strengths alone, we express the action of $\E$ in the $LC_2$ formalism where all unphysical degrees of freedom have been eliminated. The explicit non-linear $\E$ action on the scalars and vector potentials to lowest non-trivial order in $\kappa$ are derived in this gauge; they stand as the starting point for our analysis.

The remaining fields of $\m N =8$ Supergravity, including the graviton, alter these transformations through Supersymmetry. The kinematical supersymmetries generate linear transformations on the chiral superfield, while the dynamical supersymmetries are non-linear. The requirement that  $\E$ commute with the kinematical light-cone Supersymmetries  yields the $\E$ transformations of all fields, including the graviton.  The order-$\kappa$ $\E$ transformations can then be  expressed in terms of transformations on the superfield. 

Extending the commutativity to the dynamical Supersymmetries enables us then to derive through algebraic consistency, the order-$\kappa^2$ form of the supersymmetry transformations.  The quartic light-cone Hamiltonian follows. 
%%%%%%%%%%%%%%%%%%%%%%%%%%%%%%%%%%%%%%%%%%%%%
%%%%%%%%%%%%%%%%%%%%%%%%%%%%%%%%%%%%%%%%%%%%%
\section{Covariant $\m N=8$ Supergravity}
$\m N=8$ Supergravity contains a graviton $h_{\mu\nu}$ and its $8$  gravitinos $\psi^i_\mu$ interacting with matter composed of 28 vectors $A^{[ij]}_\mu$, 56 spinors $\chi^{[ijk]}$, and 70 scalars $C^{[ijkl]}$, labelled with $SO(8)$ indices, $i,j,k,l=1,2,\dots, 8$.  The much larger Cremmer-Julia $\E$ symmetry  acts on the scalars and the field strengths, and we begin with the manifestly $SO(8)$ symmetric order-$\kappa^2$ $\m N=8$ Supergravity Lagrangian \cite{deWit:1977fk} with those fields only.  The scalar part is given by
 
 \bea\label{scalarLagr}
{\cal L}_{S} ~=~-\,\frac{1}{48}\left\{\,
\d_{\mu} C^{ijkl}\, \d^{\mu}\overline{C}^{ijkl} 
+\frac{\kappa^2}{2}\, C^{ijkl}\,\overline{C}^{klmn}\,\d_{\mu} C^{mnpq}\, \d^{\mu}\overline{C}^{pqij} + {\cal O}(\kappa^3) \right\},
\eea
where the scalar fields satisfy

\bea\label{selfdualscalar}
C^{\,ijkl}_{}~=~\frac{1}{4!} \,\epsilon^{ijklmnpq}\,{\overline C}^{\,mnpq} \ .
\eea
The Lagrangian with the field strengths is given by
 
\begin{eqnarray}\label{FGplus}
\mathcal L^{}_{V}
~=-\frac{1}{8}\,{\m F}^{\,ij}_{\mu\nu}\,{\m G}^{\mu\nu\,ij}_{}+~c.c.\ .
\end{eqnarray} 
written in terms of the self-dual complex field strengths  

\begin{equation}\label{FandFbar}
{\m F}^{\mu\nu\,ij}_{}~=~\frac{1}{2} F^{\mu\nu\,ij}_{}+ \frac{i}{2}\widetilde F^{\mu\nu\,ij}_{}\ ,\end{equation}
and 

\bea
\m{G}^{\mu\nu\,ij}~ =~{\m F}^{\mu\nu\,ij}_{}+\kappa \,\overline C^{ijkl}_{}{\m F}^{\mu\nu\,kl}_{}+ \frac{\kappa^2}{2}
\overline{C}^{ijkl} \overline{C}^{klmn}\,\m{F}^{\mu\nu\,mn}\,+\,\m O(\ka^3)\ ,
\eea
is linear in the field strengths.

%%%%%%%%%%%%%%%%%%%%%%%%%%%%%%%
%%%%%%%%%%%%%%%%%%%%%%%%%%%%%%
\subsection{$SU(8)$ and $\E$ Dualities}
The electro-magnetic duality transformations exchange equations of motion

\bea\label{eom}
 \partial^{}_\mu\left({\m G}^{\mu\nu\,ij}_{}+\overline{\m G}^{\mu\nu\,ij}_{}\right)~=~0\ ,
\eea
for Bianchi identities

\bea\label{Bianchi}
\partial^{}_\mu\left(\m F^{\mu\nu\,ij}_{}-\overline{\m F}^{\mu\nu\,ij}_{}\right)~=~0\ .
\eea
These equations are manifestly $SO(8)$ covariant. We can elevate this symmetry to  
$SU(8)$ \cite{deWit:1978sh} on the complex field strengths by demanding
 
\begin{equation}\label{}
\delta\, \left({\m G}^{\mu\nu\,ij}_{}+{\m F}^{\mu\nu\,ij}_{}\right)~=~\left(R^{ik}_{}+iS^{ik}_{}\right)\left({\m G}^{\mu\nu\,kj}_{}+{\m F}^{\mu\nu\,kj}_{}\right)-(i~\leftrightarrow~ j)\ ,
\end{equation}
transforming as $\bf{28}$,
while  the other combinations $({\m G}^{\mu\nu\,ij}_{}-{\m F}^{\mu\nu\,ij}_{})$  transform  as the complex conjugate $\bf \overline{28}$,

\begin{equation}\label{}
\delta\, \left({\m G}^{\mu\nu\,ij}_{}-{\m F}^{\mu\nu\,ij}_{}\right)~=~\left(R^{ik}_{}-iS^{ik}_{}\right)\left({\m G}^{\mu\nu\,kj}_{}-{\m F}^{\mu\nu\,kj}_{}\right)-(i~\leftrightarrow~ j)\ .
\end{equation}
where  $R^{ij}$ are the $28$ {\em real} antisymmetric rotation tensors which generate $SO(8)$, and $S^{ij}$ are  $35$ {\em real} symmetric traceless matrices in  the coset $SU(8)/SO(8)$.  The $SU(8)/SO(8)$ coset transformations $\delta'$ on the complex field strengths  

\begin{equation}\label{FandG}
\delta'\, {\m F}^{\mu\nu\,ij}_{}~=~iS^{ik}_{}{\m G}^{\mu\nu\,kj}_{}-(i~\leftrightarrow~ j)\ ,\qquad \delta'\, {\m G}^{\mu\nu\,ij}_{}~=~iS^{ik}_{}{\m F}^{\mu\nu\,kj}_{}-(i~\leftrightarrow~ j)\ ,
\end{equation}
are the duality transformations which map the equations of motion into the Bianchi identities and {\em vice-versa}

\begin{equation}\label{}
\delta' \left\{ \d_\mu ({\m G}^{\mu\nu\,ij}_{}+\overline{\m G}^{\mu\nu\,ij}_{})\right\} ~=~
i\, S^{ik} \, \d_\mu ({\m F}^{\mu\nu\,kj}_{}-\overline{\m F}^{\mu\nu\,kj}_{})~-~(i \leftrightarrow j) \ .
\end{equation}
The $SU(8)/SO(8)$ transformations are only symmetries of the equations of motion and the Bianchi identities, but not of the Lagrangian.

Consistency of the coset variation of this expression with the two variations of (\ref{FandG}) requires that the scalar fields transform linearly under the full $SU(8)$, that is  

\begin{eqnarray}\label{nonlinearC}
\delta'\,\overline C^{ijkl}_{}&=&-\,i\,S^{im}_{}\,\overline C^{mjkl}_{}\,-\, (\,i\leftrightarrow\,j\,)\,
-\, (\,i\leftrightarrow\,k\,)\,
-\, (\,i\leftrightarrow\,l\,)\ ,
\end{eqnarray}
i.e. as a $ \bf{70}$.
This is an exact equation with no order $\kappa$ corrections. It follows that the scalar Lagrangian (\ref{scalarLagr}) is $SU(8)$ invariant. On the other hand, the complex field strengths have more complicated non-linear coset transformation  

\begin{eqnarray}\label{nonlinearF}
\delta'\,{\m F}^{\mu\nu\,ij}_{}&=&i\,S^{im}_{}\left({\m F}^{\mu\nu\,mj}_{}+\kappa\overline C^{mjkl}_{}{\m F}^{\mu\nu\,kl}_{}\,+\, \frac{\kappa^2}{2}
\overline{C}^{mjkl} \overline{C}^{klpq}\,\m{F}^{\mu\nu\,pq}+ \m O(\ka^3)\right) \nn\\
&&~-~(i\leftrightarrow j)\ .
\end{eqnarray}
The terms on the right-hand-side transform differently order by order in $\kappa$: $\m{F}^{\mu\nu\,mj}\sim\bf 28$, while $\overline{C}^{\,mjkl}\m{F}^{\mu\nu\,kl}\sim \overline{\bf 28}$, and the order $\kappa^2$ term has even more complicated coset transformations. Yet, one can check that the commutator of two such variations closes on $SO(8)$ transformation, as required. The extension to $SU(8)$ duality on the field strengths is meaningful only in the interacting case when $\ka\,\ne\,0$, since    $\m{G}^{\mu\nu\, ij} -\m{F}^{\mu\nu\, ij}={\cal O}(\kappa)$. 

%%%%%%%%%%%%%%%%%%%%%%%%%
Cremmer and Julia extended the duality symmetries to the non-compact $\E$ \cite{Cremmer:1979up}.  Assemble the complex field strengths in one column vector with $56$ complex entries  \cite{de Wit:1982ig}

\begin{equation}\label{XY}
Z^{\mu\nu}_{}~=~\begin{pmatrix} {\m G}^{\mu\nu\,ij}_{}+{\m F}^{\mu\nu\,ij}_{} \\ {\m G}^{\mu\nu\,ij}_{}-{\m F}^{\mu\nu\,ij}_{}\end{pmatrix}~\equiv~\begin{pmatrix}X^{\mu\nu\,ab}_{}\\ Y^{\mu\nu}_{~~~\,ab}\end{pmatrix}\ ,
\end{equation}
where $a,\, b$ are $SU(8)$ indices, with upper(lower) antisymmetric indices for $\bf 28(\bf \overline{28})$. 
Its two components 

\begin{eqnarray}
X^{\mu\nu\,ab}_{}&=&2{\m F}^{\mu\nu\,ij}_{}+\kappa \overline C^{ijkl}_{}{\m F}^{\mu\nu\,kl}_{} + \frac{\kappa^2}{2}
\overline{C}^{ijkl} \overline{C}^{klmn}\,\m{F}^{\mu\nu\,mn} \,+\,\m O(\ka^3)\ ,\label{k2X}\qquad 
\\
Y^{\mu\nu}{}_{\,ab}&=&
\kappa \overline C^{ijkl}_{}{\m F}^{\mu\nu\,kl}_{} + \frac{\kappa^2}{2}
\overline{C}^{ijkl} \overline{C}^{klmn}\,\m{F}^{\mu\nu\,mn} \,+\,\m O(\ka^3) \ ,\label{k2Y} 
\end{eqnarray}
are {\em not independent}, but related by 

\begin{equation}\label{YXrelated}
Y^{\mu\nu}{}_{ab}\,-\,\frac{\kappa}{2}\,\overline C_{abcd}\,X^{\mu\nu\,cd}_{}\,+\,\m O(\kappa^2)~=~0\ .
\end{equation}
The equations of motion (\ref{eom}) and  Bianchi identities (\ref{Bianchi}) can be written in terms of $Z^{\mu\nu}$ 

\begin{equation}\label{ZZ}
\d_\mu\, \left( Z^{\mu\nu} \,+\,\widetilde{Z}^{\mu\nu}\right)~=~0\ ,
\end{equation}
where 

\[
\widetilde{Z}^{\mu\nu}~\equiv~ \begin{pmatrix}0&\bf 1\\ \bf  1&0\end{pmatrix}\, \overline{Z}^{\mu\nu}=~\begin{pmatrix} \overline{\m G}^{\mu\nu\,ij}_{}-\overline{\m F}^{\mu\nu\,ij}_{} \\ \overline{\m  G}^{\mu\nu\,ij}_{}+\overline{\m F}^{\mu\nu\,ij}_{}\end{pmatrix}~=~\begin{pmatrix}\overline Y^{\mu\nu}{}^{\,ab}\\ \overline X^{\mu\nu}{}_{\,ab}\end{pmatrix} \ .
\]
The upper component of (\ref{ZZ}) is the sum of the equations of motion and the Bianchi identities, and the lower component the difference. It follows that the duality transformations are those which act the same way on both $Z^{\mu\nu}$ and $\widetilde{Z}^{\mu\nu}$. Explicitly, under the coset transformation denoted by $\deltab$

\begin{eqnarray}
\deltab\, X^{\mu\nu\,ab}&=&\Xi^{abcd}\,Y^{\mu\nu}{}_{cd} \label{XNNX}  \ ,\\
\deltab\, Y^{\mu\nu}{\,}_{ab}&=&\overline{\Xi}_{abcd}\,X^{\mu\nu\,cd}\label{YNNY}\  ,
\end{eqnarray}
transform $\bf 28$ into $\overline{\bf 28}$ and vice versa. It can be checked that such transformations with  real  $\Xi^{abcd}$ leave both equations of motion and Bianchi identities invariant, while those with pure imaginary $\Xi^{abcd}$  are duality transformations which interchange the two. The transformations must  respect the constraint (\ref{YXrelated}) between the upper and lower components of $Z^{\mu\nu}$ 

\[
\deltab Y^{\mu\nu}{}_{ab}~=~ \frac{\ka}{2} \deltab \left( \overline{C}_{abcd}\, X^{\mu\nu\,ab}\right)\,+\,\m O(\ka^2)\ ,
\]
that is 

\[
\overline{\Xi}_{abcd} X^{\mu\nu\,cd}~=~
\frac{\ka}{2}\, \deltab \overline{C}_{abcd} X^{\mu\nu\,cd}\,+\,\frac{\ka}{2}\overline{C}_{abef} \Xi^{efmn}(\frac{\ka}{2}\overline{C}_{mncd}X^{\mu\nu\,cd})\,+\,\m O(\ka^2)\ .
\]
It follows that the scalars must transform non-linearly  as

\begin{eqnarray}\label{scalarsu8-1}
\deltab \overline{C}_{abcd} ~=~ \frac{2}{\ka} \overline{\Xi}_{abcd} -\frac{\ka}{2}\overline{C}_{ef[ab}\, \overline{C}_{cd]mn}\Xi^{efmn}\, +\m O(\kappa^3)\ ,
\end{eqnarray}
where the indices inside the square brackets are antisymmetrized. Since the scalars satisfy the self duality condition (\ref{selfdualscalar}), so must  $\Xi^{abcd}$ 

\begin{equation}\label{Nduality}
\Xi^{abcd}~=~\frac{1}{4!}\,\epsilon^{abcdefgh}\,\overline{\Xi}_{efgh}\ ,
\end{equation}
which restricts $\Xi^{abcd}$ to $70$ real parameters. It also means that the extra term in (\ref{scalarsu8-1}) is self-dual. Repeated use of (\ref {YNNY}) yields the commutator 

\[
\left[\, \deltab_1, \,\deltab_2 \right] X^{\mu\nu\,ab}~=~
\left( \Xi_{(2)}^{abef}\,\overline{\Xi}^{}_{(1)}{}_{efcd} \, - \,  \Xi_{(1)}^{abef}\,\overline{\Xi}^{}_{(2)\,efcd}\right) X^{\mu\nu\,cd}\ .
\]
We can show \cite{BKR} that the duality requirement  ({\ref {Nduality}}) on the parameters of this commutator yields exactly the $63$ parameters of $SU(8)$, resulting in a 133-parameter group, the non-compact $\E$ since the $\E/SU(8)$ transformations are not unitary.  The $\E/SU(8)$ transformations of the complex field strengths follow

\begin{eqnarray}\label{deltaFto2}
\deltab\,{\m F}^{\mu\nu\,ij} & =& -\,\overline \Xi^{ijkl} {\m F}^{\mu\nu\,kl} 
+\frac{\kappa}{2} \left(  \Xi^{ijkl} - \overline{ \Xi}^{ijkl} \right)\overline{C}^{klmn}  {\m F}^{\mu\nu\,mn}\nn\\&& +\,\frac{\kappa^2}{4} \left(  \Xi^{ijkl} - \overline{ \Xi}^{ijkl} \right)\overline{C}^{klmn} \overline{C}^{mnpq}  {\m F}^{\mu\nu\,pq}\,+\,\m O(\ka^3)\ .
\end{eqnarray}
As we mentioned before, this equation is meaningful only when $\ka\ne0$. While the scalar part of the Lagrangian $\m L_S$ is $\E$-invariant, the vector Lagrangian $\m L_V$ is not. Invariance is attained only after invoking the equations of motion. 

%%%%%%%%%%%%%%%%%%%%%%%%%%%%%%%%%%%%%%%%%%%%%
%%%%%%%%%%%%%%%%%%%%%%%%%%%%%%%%%%%%%%%%%%%%%
\section{$E_{7(7)}$ Invariance on the Light-Cone}
The Abelian field strengths are written in terms of the potentials $A^{ij}_\mu$ through

\[
F_{}^{\mu\nu\,ij}~=~\partial_{}^\mu A^{\nu\,ij}_{}-\partial_{}^\nu A^{\mu\,ij}_{}\ .
\]
In the  $LC_2$ formalism we choose the gauge conditions 

\begin{equation}\label{LGgauge}
A^{+\,ij}~=~\frac{1}{\sqrt{2}}\left( A^0 + A^3\right)^{ij}~=~0\ ,
\end{equation}
{\em and}  invert the equations of motion to express $A^{-\,ij}$ in terms of the remaining variables in the theory, the physical  transverse components of the {\em complex} vector potentials

\[
\bar{A}^{ij}~=~ \frac{1}{\sqrt{2}}(A^1 \,+\, i\, A^2)^{ij}\,;\qquad {A}^{ij}~=~ \frac{1}{\sqrt{2}}(A^1 \,-\, i\, A^2)^{ij}\ .
\]
A lengthy but straightforward computation yields
 
\begin{eqnarray}\label{eqmotionk2}
A^{-\,ij}_{}&\equiv&\frac{1}{\sqrt{2}}\left( A^0 - A^3\right)^{ij}\nn \\ 
&=&\frac{\d}{\d^+}\,A^{ij}+\frac{\overline\partial}{\d^+}\,\bar A^{ij}_{}-\ka\frac{1}{\d^+}\left(\overline C^{ijkl}_{}\partial\,A^{kl}_{}\right)-\ka\frac{1}{\d^{+}}\left(  C^{ijkl}_{}\overline \partial\,\bar A^{kl}_{}\right) \nonumber\\
&&+\,\kappa\frac{\partial}{\partial^{+2}}\,(\, \overline C^{ijkl}_{}\partial^+\,A^{kl}_{}\,)\,+\kappa\frac{\overline\partial}{\partial^{+2}}(\, C^{ijkl}_{}\partial^+\,\bar A^{kl}_{}\,)\nn\\
&&+\,\frac{\kappa^2}{2}\,\frac{1}{\d^+} \left[ \,C^{ijkl}\overline{C}^{klmn}\d A^{mn}+\overline{C}^{ijkl}{C}^{klmn} \bar \d \bar A^{mn}    \right. \nn \\
&&\left. -( C^{ijkl}+\overline{C}^{ijkl}) \frac{\d}{\d^+}(\overline{C}^{klmn}\d^+ A^{mn})  -(C^{ijkl}+\overline{C}^{ijkl}) \frac{\bar\d}{\d^+}(C^{klmn}\d^+ \bar A^{mn})\nn\right.\\
&& \left. + \frac{\d}{\d^+}\,( \overline{C}^{ijkl}\overline{C}^{klmn}\d^+  A^{mn}\,) + \frac{\bar\d}{\d^+}(\,C^{ijkl} C^{klmn}\d^+  \bar A^{mn}) \right] 
+\, \m O(\kappa^3)\ ,
\end{eqnarray}
where 

\[
{\bar\partial} =\frac{1}{\sqrt 2}\,(\,{\partial_1}\,-\,i\,{\partial_2}\,)\ , \qquad {\partial} =\frac{1}{\sqrt 2}\,(\,{\partial_1}\,+\,i\,{\partial_2})\ ,\qquad \partial^+=\frac{1}{\sqrt 2}\,(\,-{\partial_0}
+{\partial_3}\,)\ 
\]
(The occurrence of the non-local operator $\frac{1}{\partial^+}$ is abundant in the $LC_2$ formalism. It is a harmless non-locality along the light-cone which is well understood). 

This enables us to find the $LC_2$ complex field strengths ${\m F}^{+- ij}$  

\begin{eqnarray}\label{boldF}
{\m F}^{+- ij} 
&=&\frac{1}{2}\left( \d^+ A^{-\,ij} + \d A^{ij} - \bar \d \bar A^{ij} \right)\nn \\
&=& \d A^{ij} \ - \frac{\kappa}{2}{\overline C}^{ijkl} \d A^{kl} -\frac{\kappa}{2}  C^{ijkl} \bar \d \bar A^{kl}  
+\frac{\kappa}{2} \frac{\d}{\d^+}({\overline C}^{ijkl} \d^+ A^{kl}) +\frac{\kappa}{2} \frac{\bar\d}{\d^+}( C^{ijkl} \d^+ \bar A^{kl}) \nn\\
&&+ \frac{\kappa^2}{4} \bigg[ C^{ijkl}\overline{C}^{klmn} \d A^{mn} +\overline{C}^{ijkl}C^{klmn}\bar \d \bar A^{mn}  \nn \\
&& \qquad- ( C^{ijkl} +\overline{C}^{ijkl} ) \left(\frac{\d}{\d^+}( \overline{C}^{klmn}\d^+ A^{mn} ) + \frac{\bar \d}{\d^+}(C^{klmn} \d^+\bar A^{mn}) \right)\nn \\
&&\qquad+\frac{\d}{\d^+} \left(\overline{C}^{ijkl}\overline{C}^{klmn} \d^+ A^{mn} \right)
+ \frac{\bar \d}{\d^+} \left(C^{ijkl}C^{klmn}\d^+ \bar A^{mn}\right)\bigg]\,
+ \cdots \ .
\end{eqnarray}
By varying this expression and using (\ref{nonlinearF}), we arrive at the non-linear transformation of the physical vector potentials under $SU(8)/SO(8)$
  
\bea\label{AAnonlinear}
\delta'\,A^{ij}_{}~=~iS^{im}_{}\left(A^{mj}_{}+\kappa\frac{1}{\partial^+}\left(\overline C^{mjkl}_{}\partial^+_{}A^{kl}_{}\right)\,+\,\m O(\kappa^{3
})\right)\,-~(i \leftrightarrow j) \ .
\eea
As in the covariant case, the terms on the right-hand-side do not share the same coset transformations.

Similarly, the coset $\E/SU(8)$ transformations of the vector potentials are obtained by substituting ${\m F}^{+- ij}$ in (\ref{deltaFto2}) with (\ref{boldF}). Remembering that the scalars transform non-linearly under $\E/SU(8)$ (\ref{scalarsu8-1}) , we find for the vector potentials

\begin{equation}\label{DA}
\deltab A^{ij}~ =~ -\overline \Xi^{ijkl}A^{kl} 
+\frac{\kappa}{2} ( \Xi^{ijkl} - \overline{\Xi}^{ijkl} )\frac{1}{\d^+} \left(\overline{C}^{klmn} \d^+ A^{mn} \right) + {\cal O}(\kappa^{3
})  \ ,
\end{equation}
which preserve helicity, and  exist as long as $\ka\ne 0$.

%%%%%%%%%%%%%%%%%%%%%%%%%%%%%%%%%%%%%%%%%%%%%
\subsection{The Vector and Scalar $LC_2$ Hamiltonians}
The vector Lagrangian (\ref{FGplus}) in the $LC_2$ gauge, obtained by setting $A^{+\,ij}=0$ and replacing $A^{-\,ij}$ using the equations of motion, is given by

\begin{eqnarray}\label{dWFgauge}
{\cal L}_V &=& \bar A^{ij}\,(\, - \d^+\,\d^- \,+\, \d\,\bar\d\,)\, A^{ij} \,+\,\frac{\kappa}{2}\bigg[\d^+A^{ij}\,\overline{C}^{ijmn}\,\d^-\bar A^{mn}\nn \\
&&+\,\d A^{ij} \left( \overline{C}^{ijkl} \d A^{kl} - \frac{\d}{\d^+}( \overline{C}^{ijkl} \d^+ A^{kl}) 
-\frac{\bar\d}{\d^+} (C^{ijkl}\d^+\bar A^{kl})\right) \, + \, c.c.
\bigg]\nn\\
&&-\,\frac{\kappa^2}{2} \frac{\bar \d}{\d^+}(\d^+\bar A^{ij} \, C^{ijkl})\, \frac{\d}{\d^+}(\overline{C}^{klmn}\d^+A^{mn}) \nn\\
&&+\,
\frac{\kappa^2}{2}\,\left[\, -\frac{1}{2}\d A^{ij} \overline{C}^{ijkl} C^{klmn} \bar \d \bar A^{mn} \,+\, 
\d A^{ij} \overline{C}^{ijkl}\,\frac{\bar \d}{\d^+}(C^{klmn}\d^+\bar A^{mn}) \,+\, c.c.\right]\nn\\
&&+\,\frac{\kappa^2}{2}\left[ \d A^{ij}\overline{C}^{ijkl} \frac{\d}{\d^+}(\overline{C}^{klmn}\d^+A^{mn}) -\frac{1}{2}\frac{\d}{\d^+}(\d^+A^{ij}\overline{C}^{ijkl}) \frac{\d}{\d^+}(\overline{C}^{klmn}\d^+A^{mn}) +\,c.c.\right]\nn\\
&&+\,\frac{\kappa^2}{4}\left[ \d^-\bar A^{ij} \overline{C}^{ijkl}\overline{C}^{klmn}\d^+A^{mn}\,-\, \frac{\d}{\d^+}(\d^+A^{ij}\overline{C}^{ijkl}\overline{C}^{klmn}) \left(\partial A^{mn}+\bar\partial\bar A^{mn}\right) +\, c.c.\right]\nn\\
&&+\,\, \m O(\ka^3)\ ,
\end{eqnarray}
while the scalar Supergravity Lagrangian (\ref{scalarLagr}) becomes

\begin{eqnarray}\label{dWFsca}
{\cal L}_{S}&=&-\frac{1}{24}
C^{ijkl}(\d^+\d^--\d\bar\d)\overline{C}^{ijkl}\nn\\ 
&&+\frac{\kappa^2}{96}\, C^{ijkl}\,\overline{C}^{klmn}\,(\d^+ C^{mnpq}\, \d^-\overline{C}^{pqij} +\d^- C^{mnpq}\, \d^+\overline{C}^{pqij}\nn \\
&&\hskip 1in-\d C^{mnpq}\,\bar  \d\overline{C}^{pqij}-\bar\d C^{mnpq}\, \d\overline{C}^{pqij})+ \m O(\ka^3)\ . \quad
\end{eqnarray}
Both contain the light-cone time derivative $\partial^-$ in their interactions. In order to have a Hamiltonian without this derivative we eliminate it by the field redefinitions

\begin{eqnarray}\label{redefined scalars}
C^{\,ijkl}&=& 
D^{\,ijkl}\,
-\,\frac{\kappa^2}{4}\, \frac{1}{\d^+}\left(D^{pq[ij}\d^+D^{kl]mn}\overline{D}_{pqmn}\right)+\nn \\
&& +\frac{3\,\kappa^2}{2\,\d^+}\left( \d^+B^{[ij}_{}\frac{1}{\d^+}(D^{kl]mn}_{}\d^+\overline B_{mn})\right)+\nn\\
&&+\frac{3\,\kappa^2}{2\cdot4!\,\d^+}\epsilon^{ijklrstu}_{}\left( \d^+\overline B_{rs}^{}\frac{1}{\d^+}(\overline D_{tumn}^{}\d^+ B^{mn})\right) \,+\, \m O(\ka^3)\ ,\quad
\end{eqnarray}

\begin{equation}\label{FR}
A^{ij}_{}~=~  B^{ij}-\frac{\kappa}{2}\frac{1}{\partial^+}\left(\overline D_{ijkl}\d^+ B^{kl}\right)
+\frac{ \kappa^2}{8}D^{ijkl}_{}\frac{1}{\d^+}\left( \d^+B^{mn}_{}\overline D_{mnkl}\right)\,+\m O(\ka^3)\ .
\end{equation}
This procedure leads to the unique Hamiltonian of the theory in component form.

The new vector potentials, $B^{ij}$ now transform {\em linearly} under $SU(8)$, \[
\delta'\,B^{ij}_{}~=~iS^{ik}_{}\,B^{kj}_{}-(i\leftrightarrow j)\ ,
\]
so that $i,j,\dots $ are now true $SU(8)$ indices; in particular, their lowering produces the ``barred" representation. The $\E/SU(8)$ variations of the redefinitions yield the transformation properties of the new fields

\bea\label{Sofar}
\deltab\,B^{ij}_{}~=~-\frac{\kappa}{4}\,\overline\Xi_{mnkl}^{}D^{ijkl}_{}B^{mn}_{}+\frac{\kappa}{4}\,\Xi^{ijkl}_{}\frac{1}{\d^+}\left(\overline D_{mnkl}^{}\d^+B^{mn}_{}\right)~+ \m O(\ka^{3
}
)\ ,
\eea

\begin{eqnarray}\label{Sofar2}
\deltab D^{ijkl}_{}&=& \frac{2}{\kappa}\,\Xi^{ijkl}_{}- \frac{\kappa}{2}\,\overline\Xi_{mnpq}^{}\frac{1}{\d^+}\left(D^{mn[kl}_{}\d^+D^{ij]pq}_{}\right)\nn\\
&&+ \,\frac{\kappa}{2}\,\Xi^{pq[ij}\frac{1}{\d^+}\left(\d^+D^{kl]mn}_{}\overline D_{pqmn}^{}\right)\nn \\
&&-\,{3\,\kappa}\left(\frac{\Xi^{mn[kl}_{}}{\d^+}\left(\d^+B^{ij]}_{}\overline B_{mn}^{}\right)+\epsilon^{ijklrstu}_{}\frac{\overline \Xi_{tumn}^{}}{4!\d^+}\left(B^{mn}_{}\d^+\overline B_{rs}^{}\right)\right)\nn \\
&&+\,\m O(\ka^{3
})\ .
\end{eqnarray}
We note that the $\E/SU(8)$  variation of the scalars  contains terms quadratic in the gauge fields. This mixing does not occur in the covariant formalism.
Complicated as they may seem, these variations are still incomplete since they do not include the other fields of the theory. We will use the Supersymmetry of $\m N=8$ Supergravity to generalize the transformations (\ref{Sofar}) and (\ref{Sofar2}) to include them, and defer to a later publication the construction of the vector and scalar Hamiltonians in component form, as well as the proof of their $\E$ invariance \cite{BKR}. 
 
%%%%%%%%%%%%%%%%%%%%%%%%%%%%%%%%%%%%%%%%%%%%
%%%%%%%%%%%%%%%%%%%%%%%%%%%%%%%%%%%%%%%%%%%%%
\section{$\m N=8$ Supergravity in Light-Cone Superspace}
The $256$ physical degrees of freedom of ${\cal N}=8$ Supergravity form {\em one} constrained chiral superfield in the superspace spanned by eight Grassmann variables, $\theta^m$ and their complex conjugates $\bar\theta_m$ $(m=1,...,8)$, on which $SU(8)$ acts linearly. 
We introduce the chiral derivatives

\bea
d^m ~\equiv~ -\frac{\d}{\d\bar\theta_m} -\frac{i}{\sqrt{2}} \theta^m\d^+ \, , ~~~ 
\bar d_m ~\equiv~ \frac{\d}{\d\theta^m} +\frac{i}{\sqrt{2}} \bar\theta_m\d^+ \ , 
\eea
which satisfy canonical anticommutation relations

\begin{equation}\label{phianticomm}
\left\{ d^m\,,\,\bar d_n\right\}~=~ -i \sqrt{2}\delta^m{}_{n} \d^+ \ .
\end{equation}
The physical degrees of freedom of  ${\cal N}=8$ Supergravity, the spin-2 graviton $h$ and $\overline h$;  eight spin-$\frac{3}{2}$ gravitinos, ${\psi}^m$ and 
${\overline \psi}_m$, twenty eight vector fields 

\[
 \overline B_{mn} \equiv \frac{1}{\sqrt{2}} \left( B^1_{mn} \,+\,i\,B^2_{mn}\right)\ ,
 \]
and their conjugates, fifty six  gauginos ${\overline \chi}_{mnp}$ and $\chi^{mnp}$, and finally seventy real scalars ${\overline D}_{mnpq}$ appear in one superfield

\begin{eqnarray}\label{superfield}
\varphi\,(\,y\,)\,&=&\,\frac{1}{{\d^+}^2}\,h\,(y)\,+\,i\,\theta^m\,\frac{1}{{\d^+}^2}\,{\overline \psi}_m\,(y)\,+\,i\,\theta^{mn}_{}\,\frac{1}{\d^+}\,{\overline B}_{mn}\,(y)\nn\  \\
\;&&-\,\theta^{mnp}_{}\,\frac{1}{\d^+}\,{\overline \chi}^{}_{mnp}\,(y)\,-\,\theta^{mnpq}_{}\,{\overline D}^{}_{mnpq}\,(y)+\,i\widetilde\theta^{}_{~mnp}\,\chi^{mnp}\,(y)\nn \\
&&+\,i\widetilde\theta^{}_{~mn}\,\d^+\,B^{mn}\,(y)+\,\widetilde\theta^{}_{~m}\,\d^+\,\psi^m_{}\,(y)+\,{4}\,\widetilde\theta\,{\d^+}^2\,{\bar h}\,(y)\ ,
\end{eqnarray}
where the bar denotes complex conjugation, and 
\[
\theta^{a_1a_2...a_n}_{}~=~\frac{1}{n!}\,\theta^{a_1}\theta^{a_2}_{}\cdots\theta^{a_n}_{}\ ,\quad \widetilde\theta^{}_{~a_1a_2...a_{n}}~=~ \epsilon^{}_{a_1a_2...a_{n}b_1b_2...b_{(8-n)}}\,\theta_{}^{b_1b_2\cdots b_{(8-n)}}\,\ .
\]
The arguments of the fields are the chiral coordinates 
\[
y~=~(x,\,\bar x,\, x^+, \,y^-\equiv x^- -\frac{i}{\sqrt2}\theta^m\bar\theta_m\, )\ ,\qquad  x=\frac{1}{\sqrt{2}}(x_1+ix_2)\ ,
\]
so that  $\varphi$ and its complex conjugate $\overline \varphi$ satisfy the chiral constraints

\bea\label{chiralconstraints}
d_{}^m\, \varphi ~=~0, \qquad \overline d^{}_m\, \overline \varphi ~=~0\ ,
\eea
The complex chiral superfield is related to its complex conjugate by the {\em inside-out constraint}

\begin{equation}\label{insideout}
\varphi~=~\frac{1}{4\,\d^{+4}}\,d_{}^1d_{}^2\cdots d_{}^8\, \overline\varphi\ ,
\end{equation}
in accordance with the duality condition of ${ D}^{mnpq}$. 

 On the light-cone, the eight kinematical supersymmetries (the spectrum-generating part of the symmetry) are linearly represented 
by the operators $q^m$  and $\bar q_m$ 

\begin{equation}
q^m ~=~ -\frac{\d}{\d\bar\theta_m} +\frac{i}{\sqrt{2}} \theta^m\d^+ \, , ~~~ 
\bar q_m ~=~ \frac{\d}{\d\theta^m} -\frac{i}{\sqrt{2}} \bar\theta_m\d^+ \ ,
\end{equation}
which also satisfy anticommutation relation

\begin{equation}
\{\, q^m\, , \, \bar q_n\,\} ~=~ i\sqrt{2}\, \delta^m{}_n \,\d^+\ ,
\end{equation}
and anticommute with the chiral derivatives. Hence, their linear action on the chiral superfield 

\bea
\delta^{}_s\,\varphi(y)~=~\overline \epsilon_m^{}\,q_{}^m\,\varphi(y)\ ,
\eea
where $\bar\epsilon_m$ is the parameter of the supersymmetry transformation, preserves chirality. The kinematical supersymmetry transformations of the physical fields are then 

\[
\delta^{}_s\,h~=~0\ ,\qquad \delta^{}_s\,\overline h~=~-i\frac{\sqrt{2}}{4}\,\overline\epsilon_m\,\psi^m_{}\ ,\]
\[
\delta^{}_s\,\psi^m_{}~=~2\sqrt{2}\,\overline\epsilon_n\partial^+\,B^{mn}_{}\ ,\qquad \delta^{}_s\,\overline\psi^{}_m~=~-\sqrt{2}\,\overline\epsilon_m\,\partial^+\,h\ ,\]
\[
\delta^{}_s\,B^{mn}_{}~=~-3i\sqrt{2}\,\overline\epsilon_p\,\chi^{mnp}_{}\ ,\qquad \delta^{}_s\,\overline B^{}_{mn}~=~-2i\sqrt{2}\,\overline\epsilon_{[m}\overline\psi^{}_{n]}\ ,\]
\[
\delta^{}_s\,\chi^{lmn}_{}~=~-\frac{\sqrt{2}}{3!}\,\overline\epsilon_k\,\partial^+D^{klmn}_{}\ ,\qquad \delta^{}_s\,\overline\chi^{}_{mnp}~=~-3\sqrt{2}\,\epsilon^{}_{[p}\,\partial^+\,\overline B^{}_{mn]}\ ,\]
and finally

\[
\delta^{}_s\,\overline D^{}_{klmn}~=~-4i\sqrt{2}\,\overline\epsilon^{}_{[n}\,\overline\chi^{}_{klm]}\ .\]

The quadratic operators 

\begin{equation}\label{su8fromq}
T^i{}_j ~=~ -\frac{i}{\sqrt{2} \,\d^+} \left( q^i \bar q_j -\frac{1}{8}\delta^i{}_j q^k \bar q_k \right) \ ,
\end{equation}
which satisfy the  $SU(8)$ algebra

\[
\  [\,T^i{}_j\,,\,T^k{}_l\,] ~=~\delta^k{}_j\, T^i{}_l - \delta^i{}_l \,T^k{}_j \ ,
\]
also act linearly on the chiral superfield 

\[
 \delta^{}_{SU_8} \, \varphi(y)~=~ \omega^j{}_{i}\,T^{i}{}_{j}\, \varphi(y)\ .
 \]

We can now include the other fields of the theory by demanding that the $\E/SU(8)$ transformations commute with the kinematical supersymmetries, that is

\bea
[\,\delta^{}_s\,,\,\deltab\,]\,\varphi(y)~=~0 \ .
\eea
We begin by applying this equation to the vector potential. By carefully choosing the parameters of both supersymmetry and of $\E/SU(8)$, we arrive at the generalization of (\ref{Sofar}) to order $\kappa$
 
\begin{eqnarray}\label{}
\boldsymbol{\delta} \,\overline B_{ij} &=&-\,\kappa\,\Xi^{klmn} \left( \frac{1}{4}\overline D_{ijkl} \, \overline B_{mn} + \frac{1}{4!} \frac{1}{\partial^+} \overline D_{klmn} \partial^+ \overline B_{ij} -
\frac{1}{4!} \epsilon_{ijklmnrs} \frac{1}{\partial^+} B^{rs} \partial^+ h \ \right.\nn\\
&&\qquad \qquad\left. +\, \frac{i}{3!}\frac{1}{\partial^+} \overline \chi_{klm} \,\overline\chi_{ijn} \,-\, \frac{i}{3!}\,\epsilon_{ijklmrst} 
\frac{1}{\partial^+}\chi^{rst}\overline\psi_n \right) \ \nn \\
&&+\, \kappa\,\overline{\Xi}_{ijkl} \,\frac{1}{\partial^+}\,\left( \frac{1}{4}\, D^{klmn}\,\partial^+ \overline B_{mn} \,-\, \frac{1}{\partial^+}\, B^{kl} \,\partial^{+2}\,h \ \right. \nn\\
&& \qquad\left.\qquad +\frac{i}{4 (3!)^2} \overline\chi_{mnp}\overline\chi_{rst} \epsilon^{klmnprst} \,-\, 3\,i\,\frac{1}{\partial^+} \chi^{kln}\partial^+ \overline\psi_n \right) \nn\\
&&+\, \m O(\ka^{3
})\ ,
\end{eqnarray}
as well as to the $\E/SU(8)$ transformations of the gravitinos since commutativity implies 

\[\delta^{}_s\,\deltab\,\overline B^{}_{ij}~=~-2i\sqrt{2}\overline\epsilon^{}_{[i}\,\deltab\, \overline\psi^{}_{j]}\nn\ .\]
The result is 

\bea\label{}
\deltab\, \overline \psi_i &=& -\,\kappa\, \Xi^{mnpq} \left( 
 \frac{1}{4!} \frac{1}{\partial^+} \overline D_{mnpq} \partial^+
\overline \psi_i + \frac{1}{3!} \overline D_{mnpi} \overline\psi_{q}
 \right.\nn\\
&& \left.-\frac{1}{4!}\epsilon_{mnpqirst} \frac{1}{\partial^+} \chi^{rst} \partial^+ h \,+\frac{1}{4} \overline\chi_{imn} \overline B_{pq} +\frac{1}{3!}\frac{1}{\partial^+} \overline \chi_{mnp} \partial^+ \overline B_{iq}\right)\nn\\
&&\,+\, \m O(\ka^{3
})\ .\quad
\eea
Applying commutativity on the gravitinos yields the $\E/SU(8)$ transformation of the graviton

\[\delta^{}_s\,\deltab\,\overline \psi^{}_{i}~=~-\sqrt{2}\,\overline\epsilon^{}_{i}\,\partial^+\,\deltab\, h\nn\ ,\]
with

\bea
\deltab\,h~=~-\kappa\,\Xi^{ijkl}_{}\left(\frac{1}{8}\overline B^{}_{ij}\overline B^{}_{kl}\,+\,\frac{1}{4!}\frac{1}{\partial^+}\overline D^{}_{ijkl}\partial^+h\,+\,\frac{i}{6}\frac{1}{\partial^+}\overline\chi^{}_{ijk}\overline\psi^{}_l\right) \,+\, \m O(\ka^{3
})\ .\quad
\eea
All these transformations are non-linear. Similar equations can be derived for the fifty six spinors and seventy scalars.

The inhomogeneous  $\E/SU(8)$ transformations  of order $\kappa^{-1}$ of the scalar fields can be expressed in superfield language, that is

\[
\deltab^{(-1)}_{}\,\varphi~=~-\frac{2}{\kappa}\,\theta^{ijkl}_{}\,\overline\Xi^{}_{ijkl}\ ,\]
which is chiral since $\E$ is a global symmetry: $\partial^+\overline\Xi_{ijkl}=0$. The order $\kappa$  transformations of the superfield itself take a particularly simple form. 
We need only require that its variation be chiral,  with the tensor structure

\[
\kappa\,\Xi^{ijkl}_{}(\cdots)^{}_{ijkl}\nn\ .\]
Assuming  that the lower indices are carried by the antichiral derivatives $\overline d_n$ leads to the unique form of the transformation to first order in $\ka$ 

$$
\frac{\kappa}{4!}\,\Xi^{ijkl}_{}\frac{1}{\partial^{+2}}\left(\overline d_{ijkl} \frac{1}{\partial^+}\varphi\,\partial^{+3}_{}\varphi \, -\,4\,\overline d_{ijk} \varphi\,\overline d_l\partial^{+2}_{}\varphi \,+\, 3\,\overline d_{ij} {\partial^+}\varphi\,\overline d_{kl}\partial^{+}_{}\varphi \right)\ ,
$$
where  $\bar d_{k..l}$ is a shorthand notation for $\bar d_k\,\cdots\,\bar d_l$. Including the inhomogeneous term, the $\E/SU(8)$ transformation can be written in a more compact way by introducing a coherent state-like representation 

\begin{equation}\label{}
\deltab\,\varphi~=~
-\frac{2}{\kappa}\,\theta^{ijkl}_{}\,\overline\Xi^{}_{ijkl}\,+\,
\frac{\kappa}{4!}\,\Xi^{ijkl}  \left(\frac{\d}{\d\eta}\right)_{ijkl}\frac{1}{\partial^{+2}}\left(e^{\eta \hat{\bar d}} \d^{+3} \varphi\, e^{-\eta \hat{\bar d}}\d^{+3} \varphi \right)\Bigg|_{\eta=0}\,+\, \m O(\ka^3)\ ,
\end{equation}
where 

$$
\eta\hat{\bar d} = \eta^m\frac{\bar d_m}{\d^+},~~{\rm and}~~\left(\frac{\d}{\d\eta}\right)_{ijkl} \equiv~ \frac{\d}{\d\eta^i}\frac{\d}{\d\eta^j}\frac{\d}{\d\eta^k}\frac{\d}{\d\eta^l}\ .
$$
We note that these $\E/SU(8)$ transformations do close properly to an $SU(8)$ transformation on the superfield

$$ [\, \deltab_1\,,\, \deltab_2\,] \, \varphi~=~ \delta_{SU(8)}\, \varphi\ .$$ 
It is chiral by construction $d^n_{}\,\deltab\varphi~=~0$, with the power of the first inverse derivative set by comparing with the graviton transformation. 
Hence, {\em all} physical fields, including the graviton transform under $\E$ and can be read off from this equation. It will be interesting to see what constraints this puts on the geometry. 

We can now extend the method to the dynamical supersymmetries, and determine the form of the interactions implied by the $\E$ symmetry.

%%%%%%%%%%%%%%%%%%%%%%%%%%%%%%%%%%%%%%%%%%%%%%
\subsection{ Superspace Action}
{The ${\cal N}=8$ Supergravity action in superspace was first obtained in \cite{Brink:1979nt} and its $LC_2$ form is derived in  \cite{Bengtsson:1983pg} to order $\kappa$,} using algebraic consistency and simplified further in \cite{Ananth:2005vg}. It is remarkably simple:

\begin{equation}\label{superfieldaction}
S = -\frac{1}{64}\int d^4x\,\int d^8\theta\,d^8\overline\theta \,\,\left\{\,-\,\overline\varphi\,\frac{\Box}{\d_{}^{+4}}\,\varphi
\,-\,2\kappa \left(\frac{1}{\d^{+2}}\overline\varphi
\,\overline\d\varphi\,\overline\d\varphi\,+\, c.c.\right)\,+\, \m O(\ka^2) \right\}\ ,
\end{equation}
where $\Box\equiv 2 \,(\,\d\bar\d\,- \,\-\d^+\d^- )$.
The light-cone superfield Hamiltonian density is then written as

\begin{eqnarray}\label{}
{\cal H}~=~2\,\overline\varphi\,\frac{\d\bar\d}{\d^{+4}}\,\varphi
\,+\,2\, \kappa \left(\frac{1}{\d^{+2}}\overline\varphi
\,\overline\d\varphi\,\overline\d\varphi\,+\, c.c.\right)\,+\, {\cal O}(\kappa^2)\ .
\end{eqnarray}
It can be derived from the action of the dynamical supersymmetries on the chiral superfield

\begin{eqnarray}\label{dyn-susy}
\delta^{dyn}_s\,\varphi&=&\delta^{dyn\,(0)}_s\,\varphi\,+\,\delta^{dyn\,(1)}_s\,\varphi\,+\,\delta^{dyn\,(2)}_s\,\varphi\,+\,\m O(\ka^3)\ ,\\
&=&\epsilon^m \left\{\frac{\partial}{\partial^+}\,\bar q_m\,\varphi+\,\kappa\,\frac{1}{\parp}\,{\Big (}\,{\bar \partial}\,{\bar d}_m\,\varphi\,{\d^{+2}}\,\varphi\,-\,\parp\,{\bar d}_m\,\varphi\,\parp\,{\bar \partial}\,\varphi\,
{\Big )}\,+\m O(\kappa^2)\right\}\ .\nn
\end{eqnarray}

We now require that the $\E/SU(8)$ commutes with the dynamical supersymmetries

\begin{equation}\label{4.18}
[\,\deltab\,,\,\delta^{dyn}_s\,]\,\varphi~=~0\ .
\end{equation}
This commutativity is valid only on the chiral superfield. 
For example, $[\,\deltab_1\,,\, \delta_{s}\,]\, \deltab_2 \varphi \,\ne\,0$, due to the non-linearity of the $\E$ transformation.
% commutativity with the supersymmetry on the chiral superfield (\ref{4.18}) , does not yield $[\,\deltab\,,\, \delta_{s}\,]\, \deltab \varphi\,=\,0 $. 
 This helps us understand how the Jacobi identity
 
$$ \left(\,  [\, \deltab_1\,,\,[\,\deltab_2\,,\, \delta_{s}\,]\,]  \,+\, [\, \deltab_2\,,\,[\,\delta_{s}\,,\, \deltab_1\,]\,] \, +\, [\, \delta_{s}\,,\,[\,\deltab_1\,,\, \deltab_2\,]\,] \,\right)\, \varphi~ =~0\ ,$$ 
is algebraically consistent. In the last  term the commutator of the  two $\E/SU(8)$ transformations,  $[\,\deltab_1\,,\, \deltab_2\,]$, yields an  $SU(8)$ under which the supersymmetry transforms. This is precisely compensated by contributions from the first two terms.

Although the dynamical supersymmetry to order $\kappa$ is already known, we re-derive $\delta^{dyn\,(1)}_s\,\varphi$ from the commutativity between the dynamical supersymmetries and $\E/SU(8)$ transformations. 

The inhomogeneous $\E$ transformations link interaction terms with different order in $\kappa$. To zeroth order, one finds

\begin{equation}\label{70swith-susy-tofirstorder}
\ [ \, \boldsymbol\delta^{(-1)}\,,\,\delta^{dyn\,(1)}_{s} \, ] \, \varphi  ~=~\deltab^{(-1)}\,\delta^{dyn\,(1)}_{s} \varphi~=~0 \ ,
\end{equation}
since $\delta^{dyn\,(1)}_{s}\,\deltab^{(-1)} \varphi~=~0$. To find $\delta^{dyn\,(1)}_{s} \varphi$ that satisfies both the above equation and the SuperPoincar\'e algebra, one may start with a general form that satisfies all the commutation relations with the kinematical SuperPoincar\'e generators (the forms of the kinematical SuperPoincar\'e generators can be found in \cite{Bengtsson:1983}),

$$ \delta^{dyn\,(1)}_{s} \varphi \propto \frac{\d}{\d a}\frac{\d}{\d b}\,\frac{1}{\d^{+(m+n+1)}} \left( e^{a \hat{\bar \d}}e^{b\,\ep \hat{\bar q}} \d^{+(2+m)}\varphi\, e^{-a \hat{\bar \d}}e^{-b\,\ep \hat{\bar q}}\d^{+(2+n)}\varphi \right) \Big|_{ a=b=0 }\ ,$$ 
where $\hat{\bar \d} = \frac{\bar\d}{\d^+}$, $\ep\hat{\bar q} = \ep^m\frac{\bar q_m}{\d^+}$.
It is not difficult to see that this form with non-negative $m,\, n$ satisfies (\ref{70swith-susy-tofirstorder}). The number of powers of $\d^+$ can be determined by checking the commutation relation between two dynamical generators $\delta_{p^-}$(Hamiltonian variation which is derived from the supersymmetry algebra) and $\delta^{}_{j^-}$(the boost which can also be obtained through $ [\,\delta^{}_{j^-}\,,\, \delta_{\bar q} \,]\,\varphi\, =\, \delta^{dyn}_{s}\varphi$), yielding that the commutator between $\delta^{}_{j^-}$ and $\delta_{p^-}$ vanishes only when $m=n=0$,
which leads to the the same form as (\ref{dyn-susy}) written in a coherent-like form

$$
\delta^{dyn\,(1)}_{s} \varphi = \frac{\ka}{2} \frac{\d}{\d a}\frac{\d}{\d b}\,\frac{1}{\d^{+}} \left[ e^{a \hat{\bar \d}}e^{b\,\ep \hat{\bar q}} \d^{+2}\varphi\, e^{-a \hat{\bar \d}}e^{-b\,\ep \hat{\bar q}}\d^{+2}\varphi \right] \Big|_{a=b=0 }\ .$$
It is worth noting that this is the solution that has the least number of powers of $\d^{+}$ in the denominator, and thus the least ``non-local''. 

The same reasoning can be applied to higher orders in $\kappa$.  To order $\ka$, we find that commutativity

$$ \ [ \, \boldsymbol\delta^{(-1)}\,,\,\delta^{dyn\,(2)}_{s} \, ] \, \varphi \, +\,[ \, \boldsymbol\delta^{(1)}\,,\,\delta^{dyn\,(0)}_{s} \, ] \, \varphi ~=~0\ $$ 
requires

\begin{eqnarray}\label{delta2qvarphi1}
&&\deltab^{(-1)}\,\delta^{dyn\,(2)}_{s} \varphi\\
&&=~\frac{\kappa}{4!} \Xi^{ijkl} \frac{1}{\partial^{+3}} \left[ \,
-\,{\bar d}^{}_{ijkl}\frac{\partial}{\partial^+} \varphi \, \partial^{+3}\epsilon\bar q \,\varphi
+\,4\, {\bar d}^{}_{ijk} \partial \varphi \, \bar d_l \partial^{+2} {\epsilon\bar q}\, \varphi 
-\, 3\,{\bar d}^{}_{ij}\partial \partial^+ \varphi\, {\bar d}^{}_{kl}\partial^+{\epsilon\bar q}\,\varphi \right.\nn\\
&&\qquad\qquad\qquad~~
~ - \,{\bar{d}}^{}_{ijkl} \frac{\epsilon\bar q}{\partial^+}\, \varphi \,\partial\partial^{+3}\varphi
+\, 4\, {\bar d}^{}_{ijk} {\epsilon\bar q}\, \varphi \, \bar d_l \partial \partial^{+2}\,\varphi  
-3 \,{\bar d}^{}_{ij} \partial^{+} {\epsilon\bar q}\varphi\, {\bar d}^{}_{kl}\partial\partial^+\varphi \nn\\
%%%
&&\qquad\qquad\qquad~~
~ +\, {\bar d}^{}_{ijkl}\frac{\partial}{\partial^{+2}} \epsilon\bar q\,\varphi \, \partial^{+4}\varphi
\,-\, 4\, {\bar d}^{}_{ijk} \frac{\partial}{\partial^+} {\epsilon\bar q} \,\varphi \, \bar d_l \partial^{+3}\,\varphi  
\,+\,  3\,{\bar d}^{}_{ij}\partial {\epsilon\bar q}\,\varphi\, {\bar d}^{}_{kl}\partial^{+2}\varphi \nn\\
%%%%
&&\qquad\qquad\qquad~~
~+\,{\bar d}^{}_{ijkl}\,\varphi \, \partial \partial^{+2}{\epsilon\bar q}\, \varphi 
\,-\,4\, {\bar d}^{}_{ijk} \partial^+\, \varphi \, \bar d_l \partial \partial^{+}{\epsilon\bar q}\,\varphi  
\,+\,  3\,{\bar d}^{}_{ij} \partial^{+2} \varphi\, {\bar d}^{}_{kl}\partial{\epsilon\bar q}\varphi \bigg]\ ,\nn
\end{eqnarray}
where $\epsilon \bar q$ denotes $\epsilon^m\bar q_m$, which can be written in a simpler form by rewriting it in terms of a coherent state-like form: 

\begin{eqnarray}\label{coherent-state-form}
&&\deltab^{(-1)}_{}\,\delta^{dyn\,(2)}_{s} \varphi \\
&&=~  
 \frac{\kappa}{2\cdot4!} \Xi^{ijkl} \frac{\d}{\d a}\frac{\d}{\d b}  
 \left(\frac{\d}{\d\eta}\right)_{ijkl}
 \frac{1}{\partial^{+3}} \left[\
 e^{a \hat{\d}} e^{b\, \epsilon\hat{\bar q}}e^{\eta\hat{\bar d}}\d^{+4} \varphi\,\, e^{-a\hat{\d}}e^{-b \,\epsilon\hat{ \bar q}} e^{-\eta\hat{\bar d}}\d^{+4} \varphi
\, \right] \Bigg|_{a=b=\eta=0}\ .\nn
\end{eqnarray}

To find $\delta^{dyn (2)}_{s}\varphi$ that satisfies (\ref{delta2qvarphi1}), consider the chiral combination

\begin{eqnarray}\label{}
Z_{mnpq}&\equiv&\left(\frac{\d}{\d\xi}\right)_{mnpq}\left( e^{\xi\hat{\bar d}}  \d^{+4}\varphi e^{-\xi\hat{\bar d}}  \d^{+4}\varphi \right) \Big|_{\xi =0}\ ,\\
&=&
\bar d^{}_{mnpq} \varphi\, \partial^{+4} \varphi \,-\,4\,\bar d^{}_{mnp}\partial^+\varphi \,\bar d_q \partial^{+3} \varphi \,+\,3\,\bar d^{}_{mn}\partial^{+2}\varphi \,\bar d^{}_{pq} \partial^{+2} \varphi \ .\nn
\end{eqnarray}
The inhomogeneous $\E$ transformation of

$$Z^{ijkl}\equiv\frac{1}{4!}\epsilon^{ijklmnpq}Z_{mnpq}\ ,$$
has the simple form 
\bea\label{}
\boldsymbol\delta^{(-1)} Z^{ijkl} =\frac{1}{4!}\epsilon^{ijklmnpq}\,\bar d^{}_{mnpq} \boldsymbol\delta^{(-1)} \varphi\, \partial^{+4} \varphi =
\frac{2}{\kappa} \,\Xi^{ijkl}\, \partial^{+4} \varphi \ ,
\eea
which leads to the solution

$$
\delta^{dyn\,(2)}_s\,\varphi~=~
\frac{\kappa^2}{2\cdot4!}\frac{\d}{\d a}\frac{\d}{\d b}\left(\frac{\d}{\d\eta}\right)_{ijkl}
\frac{1}{\d^{+4}}\left(\,
 e^{a \hat{\d}+b\, \epsilon\hat{\bar q}+\eta\hat{\bar d}}\d^{+5} \varphi\,\,
 e^{-a \hat{\d}-b\,\epsilon \hat{\bar q}-\eta\hat{\bar d}}  Z^{ijkl}
\right)
\, \Bigg|_{a=b=\eta=0}\ ,
$$
where we have fixed the ambiguity discussed earlier by choosing the expression with the least number of $\d^+$ in the denominator. Its algebraic consistency should be checked in a future publication.  This coherent state-like form is very efficient; Written out explicitly $\delta^{dyn\,(2)}_s\varphi$ consists of 60 terms.

The dynamical supersymmetry is then written in terms of the coherent state-like form, 

\begin{eqnarray}\label{solq2}
\delta^{dyn}_s\,\varphi
&=& \frac{\d}{\d a}\frac{\d}{\d b} \Bigg\{ e^{a\hat{\d}} e^{b\,\ep \hat{\bar q}} \d^+\varphi \, +\,
\frac{\kappa}{2}\,\frac{1}{\d^{+}}\left(e^{a\hat{\bar \d}+b\,\ep\hat{\bar q}} \d^{+2}\varphi e^{-a\hat{\bar \d}-b\,\ep\hat{\bar q}} \d^{+2}\varphi \right) \nn\\
&&~+\,\frac{\kappa^2}{2\cdot4!}\left(\frac{\d}{\d\eta}\right)_{ijkl}
\frac{1}{\d^{+4}}\left(\,
 e^{a \hat{\d}+\, b\,\epsilon\hat{\bar q}+\eta\hat{\bar d}}\d^{+5} \varphi\,\,
 e^{-a \hat{\d}-\,b\,\epsilon \hat{\bar q}-\eta\hat{\bar d}}  Z^{ijkl}
\right)  \nn \\
&&~+\,\m O(\ka^3) \Bigg\}
 \Bigg|_{a=b=\eta=0} \ .
\end{eqnarray}

%%%%
We now use the fact, as  Ananth et al \cite{Ananth:2006fh} have shown, that the $\m N =8$ supergravity light-cone Hamiltonian can be written as a quadratic form (to order $\kappa^2$), 

\[ \m 
H~= ~ \frac{1}{4\sqrt2}\, \left( \m W_m\,,\, \m W_m\right)~\equiv~\frac{2\,i}{4\sqrt2} \int  d^8 \theta\,d^8\bar\theta\,d^4x\,\,  \overline{\m W}_m\frac{1}{\d^{+3}}\m W_m \ ,  \]
where the fermionic superfield $\m W_m$ is  the dynamical supersymmetry variation of $\varphi$ 

\[
\delta^{dyn}_{s} \varphi~\equiv~ \epsilon^m_{}\,\m W^{}_m\ ,\]
with 

\[\m W_m~=~ \m W^{(0)}_{m} \, +\,\,\m W^{(1)}_{m} \, +\,\,\m W^{(2)}_{m} +\,\cdots\ . \]
Up to order $\ka$, the Hamiltonian is simply

\begin{equation}\label{}
\m H~=~ \frac{1}{4\sqrt2}\, \left[ \left( \m W^{(0)}_m\,,\, \m W^{(0)}_m\right)\,+ \,\left( \m W^{(0)}_m\,,\, \m W^{(1)}_m\right)\,+ \, \left( \m W^{(1)}_m\,,\, \m W^{(0)}_m\right)\right] \ ,
\end{equation}
while the Hamiltonian of order $\ka^2$ consists of three parts:

\begin{equation}\label{}
 \m H^{\ka^2} ~=~ \frac{1}{4\sqrt2}\, \left[  \left( \m W^{(1)}_m\,,\, \m W^{(1)}_m\right)\,+\,\left( \m W^{(0)}_m\,,\, \m W^{(2)}_m\right)\,+ \, \left( \m W^{(2)}_m\,,\, \m W^{(0)}_m\right)\right] \ , 
\end{equation}
where the first part was computed by Ananth et al \cite{Ananth:2006fh}

\begin{eqnarray}\label{}
&&\left( \m W^{(1)}_m\,,\, \m W^{(1)}_m\right)~=~ i\frac{\ka^2}{2}\frac{\d}{\d a}\frac{\d}{\d b}\frac{\d}{\d r}\frac{\d}{\d s} \int d^8 \theta\,d^8\bar\theta\,d^4x\,\,\\
&&
\frac{1}{\d^{+5}} \left(e^{a\hat{\d}+b\hat{q}^m} \d^{+2}\overline\varphi e^{-a\hat{\d}-b\hat{q}^m} \d^{+2}\overline\varphi \right)
\left(e^{r\hat{\bar \d}+s\hat{\bar q}_m} \d^{+2}\varphi e^{-r\hat{\bar \d}-s\hat{\bar q}_m} \d^{+2}\varphi \right)\Big|_{a=b=r=s=0}\ ,\qquad \nn
\end{eqnarray}
 and the second and  third parts are complex conjugate of each other.  It suffices to consider  

\begin{eqnarray}\label{}
 \left( \m W^{(0)}_m\,,\, \m W^{(2)}_m\right)&=& i\frac{\ka^2}{4!}\frac{\d}{\d a}\frac{\d}{\d b}\left(\frac{\d}{\d\eta}\right)_{ijkl}\int d^8 \theta\,d^8\bar\theta\,d^4x\,\,\\
&&~\frac{\bar \d}{\d^+} q^m \overline \varphi\, 
\frac{1}{\partial^{+7}}\left(\,
 e^{a \hat{\d}+b\, \hat{\bar q}_m+\eta\hat{\bar d}}\d^{+5} \varphi\,\,
 e^{-a \hat{\d}-b\, \hat{\bar q}_m-\eta\hat{\bar d}}  Z^{ijkl}
\right) \Bigg|_{a=b=\eta=0} \ .\nn
\end{eqnarray}
Integration by parts with respect to $\bar d$'s and use of the inside-out constraint (\ref{insideout}) allow for  an efficient rearrangement of terms to yield the final expression

\begin{eqnarray}\label{}
&&\left( \m W^{(0)}_m\,,\, \m W^{(2)}_m\right) \\
&&=~-i
\frac{\ka^2}{4!} \frac{\d}{\d a}\frac{\d}{\d b} \int d^8 \theta\,d^8\bar\theta\,d^4x\,\frac{\bar \d}{\d^{+4}} q^m {d}^{ijkl}\overline \varphi
\left(  e^{a \hat{\d}\, +\, b\, \hat{\bar q}_m} \d^+\overline\varphi \,\,e^{-a \hat{\d}\, -\, b \hat{\bar q}_m}
\frac{1}{\d^{+4}} Z_{ijkl}
\right)\Bigg|_{a=b=0} \ .\nn
\end{eqnarray}
Therefore, the Hamiltonian to order $\kappa^2$  is written as

\begin{eqnarray}\label{}
 &&\m H^{\ka^2} ~=~i\,\frac{\ka^2}{4\sqrt2} \int d^8 \theta\,d^8\bar\theta\,d^4x\,\,\frac{\d}{\d a}\frac{\d}{\d b}\\
&&\quad\Bigg\{\frac{1}{2} \frac{\d}{\d r}\frac{\d}{\d s} \frac{1}{\d^{+5}} \left(e^{a\hat{\d}+b\hat{q}} \d^{+2}\overline\varphi e^{-a\hat{\d}-b\hat{q}} \d^{+2}\overline\varphi \right)\left(e^{r\hat{\bar \d}+s\hat{\bar q}} \d^{+2}\varphi e^{-r\hat{\bar \d}-s\hat{\bar q}} \d^{+2}\varphi \right)\nn\\
&&~~~~-\, \left[\,\frac{1}{4!}\, \frac{\bar \d}{\d^{+4}} q^m {d}^{ijkl}\overline \varphi
\left(  e^{a \hat{\d}\, +\, b\, \hat{\bar q}_m} \d^+\overline\varphi \,\,e^{-\,a\, \hat{\d} \,-\, b\, \hat{\bar q}_m}
\frac{1}{\d^{+4}} Z_{ijkl}
\right) +c.c. \right] \Bigg\} \Bigg|_{a=b=r=s=0} \ ,\nn
\end{eqnarray}
to be compared with the 96 terms of Ananth et al \cite{Ananth:2006fh}! 

\section{Conclusions and Outlook}

In this paper, we have explicitly derived the non-linear $\E$ transformations on the $256$ physical fields of $\m N=8$ Supergravity. We found that they can be elegantly written in terms of the constrained chiral superfield in light-cone superspace, at least to order $\kappa$. In this gauge, all fields, including the graviton transform under $\E$. This is to be compared with the original covariant formulation of Cremmer and Julia in which the graviton is invariant. 
The process of gauge fixing, elimination of the unphysical degrees of freedom, and the preservation of kinematical supersymmetry, requires that all fields transform. 
{We note that the form of the Lagrangian in the $LC_2$ formulation that involves no time derivatives in the interaction is unique. Since the kinetic term contains the time derivatives, any non-linear field redefinition would reintroduce time derivatives in the interaction terms. Hence there is no possibility to find a change of field degrees of freedom which could lead to a graviton field invariant under $E_{7(7)}$.}
The simplicity of our result lends hope for the existence of a compact all-orders in $\kappa$ formulation of these non-linear $E_{7(7)}$ transformations\footnote{
As this paper was being revised, a preprint appeared \cite{Kallosh:2008ic} which provided the action of  the $E_{7(7)}$ generators to all orders in $\kappa$. We note that in their analysis, performed in the covariant formalism, only the vectors and scalars transform under the coset $E_{7(7)}/SU(8)$, but in the $LC_2$ formalism we use all physical fields transform. Therefore the all-order expressions for the coset transformation on fields are different from those in the covariant formalism.}  (as in non-linear $\sigma$-models).

In light-cone superspace, the symmetries of the theory can be identified with the semi-direct product of the $\m N=8$ SuperPoincar\' e group with $\E$. Its non-linearity indicates that while the eight supercharges do transform under its compact subgroup, they are invariant under the $\E/SU(8)$ coset. These symmetries can now be used to find the dynamical supersymmetry, and then the light-cone Hamiltonian. We use {\em both} $\E$ and the superPoincar\'e algebra to find the dynamical supersymmetry transformations of the superfield to order $\kappa^2$. In the superfield language, it is also a remarkably simple expression, which suggests that an all-order in $\kappa$ expression may be feasible. The light-cone Hamiltonian that follows is equally simple.  
 
Our results indicate that the light-cone superspace formalism, although awkward for many detailed calculations, produces tractable results that are bound to shed further light on the structure of $\m N =8$ Supergravity.

The formalism is also suitable to examine how general these symmetries are, such as if they can be extended to other dimensions. All these questions will be discussed and more detailed proofs of the formulae in this paper in forthcoming papers \cite{BKR}.
\\

\begin{flushleft}  
{\Large \bf Acknowledgements}
\end{flushleft}
LB would like to thank Prof. Edward Witten for an invitation to the Institute of Advanced Study where part of this work was done, as well as the Institute for Fundamental Theory at the University of Florida. He also thanks Prof. Hermann Nicolai for useful discussions.  PR thanks  the Institute of Advanced Study and the Ambrose Monell foundation for their support. 
SK thanks Prof. Charles B. Thorn for useful discussions, and is supported by a McLaughlin Dissertation Fellowship from the University of Florida. Two of us (PR and SK) are also supported by the Department of Energy Grant No. DE-FG02-97ER41029.


\begin{thebibliography}{Ref}

\bibitem{Cremmer:1979up}
  E.~Cremmer and B.~Julia,
  ``The N=8 Supergravity Theory. 1. The Lagrangian,''
  Phys.\ Lett.\  B {\bf 80}, 48 (1978); 
%  E.~Cremmer and B.~Julia,
``The SO(8) Supergravity,''
  Nucl.\ Phys.\  B {\bf 159}, 141 (1979).
  
  \bibitem{BSS}
L.~Brink, J.~H.~Schwarz and J.~Scherk,
``Supersymmetric Yang-Mills Theories,''
Nucl.\ Phys.\ B {\bf 121}, 77 (1977); F.~Gliozzi, J.~Scherk and D.~I.~Olive,
``Supersymmetry, Supergravity Theories And The Dual Spinor Model,''
Nucl.\ Phys.\ B {\bf 122}, 253 (1977).

%\cite{Brink:1982pd}
\bibitem{Brink:1982pd}
  L.~Brink, O.~Lindgren and B.~E.~W.~Nilsson,
  ``N=4 Yang-Mills Theory On The Light Cone,''
  Nucl.\ Phys.\  B {\bf 212}, 401 (1983).
  %%CITATION = NUPHA,B212,401;%%
  
%\cite{Berends:1988zp}
\bibitem{Berends:1988zp}
  F.~A.~Berends, W.~T.~Giele and H.~Kuijf,
  ``On relations between multi - gluon and multigraviton scattering,''
  Phys.\ Lett.\  B {\bf 211}, 91 (1988).
  %%CITATION = PHLTA,B211,91;%%
  
%\cite{Kawai:1985xq}
\bibitem{Kawai:1985xq}
  H.~Kawai, D.~C.~Lewellen and S.~H.~H.~Tye,
  ``A Relation Between Tree Amplitudes Of Closed And Open Strings,''
  Nucl.\ Phys.\  B {\bf 269}, 1 (1986).
  %%CITATION = NUPHA,B269,1;%%
  
   
\bibitem{Ananth:2005zd}
  S.~Ananth, L.~Brink, S.~-S.~Kim and P.~Ramond,
  ``Non-linear realization of $PSU(2,2|4)$ on the light-cone,''
  Nucl.\ Phys.\  B {\bf 722}, 166 (2005)
%  [arXiv:hep-th/0505234].

%\cite{Ananth:2006fh}
\bibitem{Ananth:2006fh}
  S.~Ananth, L.~Brink, R.~Heise and H.~G.~Svendsen,
  ``The N=8 Supergravity Hamiltonian as a Quadratic Form,''
  Nucl.\ Phys.\  B {\bf 753}, 195 (2006)
  % [arXiv:hep-th/0607019].

\bibitem{UVfinite}
  S.~Mandelstam,
  ``Light Cone Superspace And The Ultraviolet Finiteness Of The N=4 Model,''
  Nucl.\ Phys.\  B {\bf 213}, 149 (1983); 
  L.~Brink, O.~Lindgren and B.~E.~W.~Nilsson,
  ``The Ultraviolet Finiteness Of The N=4 Yang-Mills Theory,''
  Phys.\ Lett.\  B {\bf 123}, 323 (1983).

 %\cite{Bern:1998ug}
\bibitem{Bern:1998ug}
  Z.~Bern, L.~J.~Dixon, D.~C.~Dunbar, M.~Perelstein and J.~S.~Rozowsky,
  ``On the relationship between Yang-Mills theory and gravity and its implication for ultraviolet divergences,''
  Nucl.\ Phys.\  B {\bf 530}, 401 (1998);
%  [arXiv:hep-th/9802162]; 
  %%CITATION = NUPHA,B530,401;%% 
  %\cite{Bern:2007hh}
%\bibitem{Bern:2007hh}
  Z.~Bern, J.~J.~Carrasco, L.~J.~Dixon, H.~Johansson, D.~A.~Kosower and R.~Roiban,
  ``Three-Loop Superfiniteness of N=8 Supergravity,''
  Phys.\ Rev.\ Lett.\  {\bf 98}, 161303 (2007)
  %[arXiv:hep-th/0702112].
  %%CITATION = PRLTA,98,161303;%%

%\cite{Green:2006gt}
\bibitem{Green:2006gt}
  M.~B.~Green, J.~G.~Russo and P.~Vanhove,
  ``Non-renormalisation conditions in type II string theory and maximal
  %supergravity,''
  JHEP {\bf 0702}, 099 (2007);
 % [arXiv:hep-th/0610299] 
  %%CITATION = JHEPA,0702,099;%%
    %\cite{Green:2006yu}
%\bibitem{Green:2006yu}
%  M.~B.~Green, J.~G.~Russo and P.~Vanhove,
  ``Ultraviolet properties of maximal supergravity,''
  Phys.\ Rev.\ Lett.\  {\bf 98}, 131602 (2007)
  %[arXiv:hep-th/0611273].
  %%CITATION = PRLTA,98,131602;%%
    
 \bibitem{BKR}
 L.~Brink, S.~-S.~Kim and P.~Ramond, in preparation. 
  
  
  %\cite{deWit:1977fk}
\bibitem{deWit:1977fk}
  B.~de Wit and D.~Z.~Freedman,
  ``On SO(8) Extended Supergravity,''
  Nucl.\ Phys.\  B {\bf 130}, 105 (1977).
  %%CITATION = NUPHA,B130,105;%%
  
%\cite{deWit:1978sh}
\bibitem{deWit:1978sh}
  B.~de Wit,
  ``Properties Of SO(8) Extended Supergravity,''
  Nucl.\ Phys.\  B {\bf 158}, 189 (1979).
  %%CITATION = NUPHA,B158,189;%%  
   
  
%\cite{de Wit:1982ig}
\bibitem{de Wit:1982ig}
  B.~de Wit and H.~Nicolai,
 ``N=8 Supergravity,''
  Nucl.\ Phys.\  B {\bf 208}, 323 (1982).


%\cite{Ananth:2004es}
\bibitem{Ananth:2004es}
  S.~Ananth, L.~Brink and P.~Ramond,
  ``Oxidizing SuperYang-Mills from (N = 4, d = 4) to (N = 1, d = 10),''
  JHEP {\bf 0407}, 082 (2004).
  % [arXiv:hep-th/0405150].
  
%\cite{Brink:1979nt}
\bibitem{Brink:1979nt}
  L.~Brink and P.~S.~Howe,
  ``The N=8 Supergravity In Superspace,''
  Phys.\ Lett.\  B {\bf 88}, 268 (1979).
  %%CITATION = PHLTA,B88,268;%%


 %\cite{Bengtsson:1983pg}
 \bibitem{Bengtsson:1983pg}
  A.~K.~H.~Bengtsson, I.~Bengtsson and L.~Brink,
  ``Cubic Interaction Terms For Arbitrarily Extended Supermultiplets,''
  Nucl.\ Phys.\  B {\bf 227}, 41 (1983).
  %%CITATION = NUPHA,B227,41;%%
 
 %\cite{Ananth:2005vg}
\bibitem{Ananth:2005vg}
  S.~Ananth, L.~Brink and P.~Ramond,
  ``Eleven-dimensional supergravity in light-cone superspace,''
  JHEP {\bf 0505}, 003 (2005).
%  [arXiv:hep-th/0501079].

\bibitem{Bengtsson:1983}
  A.~K.~H.~Bengtsson, I.~Bengtsson and L.~Brink,
  ``Cubic Interaction Terms For Arbitrary Spin,''
  Nucl.\ Phys.\  B {\bf 227}, 31 (1983);
  %%CITATION = NUPHA,B227,31;%%
  A.~K.~H.~Bengtsson, I.~Bengtsson and L.~Brink,
  ``Cubic Interaction Terms For Arbitrarily Extended Supermultiplets,''
  Nucl.\ Phys.\  B {\bf 227}, 41 (1983).
  %%CITATION = NUPHA,B227,41;%%
  
%\cite{Kallosh:2008ic}
\bibitem{Kallosh:2008ic}
  R.~Kallosh and M.~Soroush,
  ``Explicit Action of E7(7) on N=8 Supergravity Fields,''
  arXiv:0802.4106 [hep-th].
  %%CITATION = ARXIV:0802.4106;%%    
\end{thebibliography}
\end{document}